\documentstyle{mn}
\newif\ifAMStwofonts
\newcommand{\lym}{Lyman-$\alpha$ }
\newcommand{\dlym}{damped Lyman-$\alpha$ }
\ifoldfss

  \ifCUPmtlplainloaded \else
    \NewTextAlphabet{textbfit} {cmbxti10} {}
    \NewTextAlphabet{textbfss} {cmssbx10} {}
    \NewMathAlphabet{mathbfit} {cmbxti10} {} 
    \NewMathAlphabet{mathbfss} {cmssbx10} {} 
  \fi
  \ifAMStwofonts
    \ifCUPmtlplainloaded \else
      \NewSymbolFont{upmath} {eurm10}
      \NewSymbolFont{AMSa} {msam10}
      \NewMathSymbol{\upi}     {0}{upmath}{19}
      \NewMathSymbol{\umu}     {0}{upmath}{16}
      \NewMathSymbol{\upartial}{0}{upmath}{40}
      \NewMathSymbol{\leqslant}{3}{AMSa}{36}
      \NewMathSymbol{\geqslant}{3}{AMSa}{3E}

      \let\geq=\geqslant 
    \fi
  \fi
\fi 
\ifnfssone
  \newmathalphabet{\mathit}
  \addtoversion{normal}{\mathit}{cmr}{m}{it}
  \addtoversion{bold}{\mathit}{cmr}{bx}{it}

  \newmathalphabet{\mathbfit} 
  \addtoversion{normal}{\mathbfit}{cmr}{bx}{it}
  \addtoversion{bold}{\mathbfit}{cmr}{bx}{it}
  \newmathalphabet{\mathbfss} 
  \addtoversion{normal}{\mathbfss}{cmss}{bx}{n}
  \addtoversion{bold}{\mathbfss}{cmss}{bx}{n}
  \ifAMStwofonts
    \ifCUPmtlplainloaded \else
      \UseAMStwoboldmath
      \makeatletter
      \new@mathgroup\upmath@group
      \define@mathgroup\mv@normal\upmath@group{eur}{m}{n}
      \define@mathgroup\mv@bold\upmath@group{eur}{b}{n}
      \edef\UPM{\hexnumber\upmath@group}
      \new@mathgroup\amsa@group
      \define@mathgroup\mv@normal\amsa@group{msa}{m}{n}
      \define@mathgroup\mv@bold\amsa@group{msa}{m}{n}
      \edef\AMSa{\hexnumber\amsa@group}
      \makeatother
      \mathchardef\upi="0\UPM19
      \mathchardef\umu="0\UPM16
      \mathchardef\upartial="0\UPM40
      \mathchardef\leqslant="3\AMSa36
      \mathchardef\geqslant="3\AMSa3E

      \let\geq=\geqslant 
    \fi
  \fi
\fi 
\ifnfsstwo

  \DeclareMathAlphabet{\mathbfit}{OT1}{cmr}{bx}{it}
  \SetMathAlphabet\mathbfit{bold}{OT1}{cmr}{bx}{it}
  \DeclareMathAlphabet{\mathbfss}{OT1}{cmss}{bx}{n}
  \SetMathAlphabet\mathbfss{bold}{OT1}{cmss}{bx}{n}
  \ifAMStwofonts
    \ifCUPmtlplainloaded \else
      \DeclareSymbolFont{UPM}{U}{eur}{m}{n}
      \SetSymbolFont{UPM}{bold}{U}{eur}{b}{n}
      \DeclareSymbolFont{AMSa}{U}{msa}{m}{n}
      \DeclareMathSymbol{\upi}{0}{UPM}{"19}
      \DeclareMathSymbol{\umu}{0}{UPM}{"16}
      \DeclareMathSymbol{\upartial}{0}{UPM}{"40}
      \DeclareMathSymbol{\leqslant}{3}{AMSa}{"36}
      \DeclareMathSymbol{\geqslant}{3}{AMSa}{"3E}

      \let\geq=\geqslant 
    
    \fi
  \fi
 
\fi 

\ifCUPmtlplainloaded \else
  \ifAMStwofonts \else 
    \def\upi{\pi}
    \def\umu{\mu}
    \def\upartial{\partial}
  \fi
\fi
\title{ORT observations of the damped Lyman-$\alpha$ system towards PKS 0201+113}
\author[Nissim Kanekar $\&$ Jayaram N Chengalur ]
       { Nissim Kanekar\thanks{nissim@ncra.tifr.res.in}
 $\&$ Jayaram N Chengalur\thanks{chengalu@ncra.tifr.res.in}
 \\
        National Centre for Radio Astrophysics, Tata Institute of Fundamental
        Research, Pune - 411007, India}
\pagerange{\pageref{firstpage}--\pageref{lastpage}}
\begin{document}
\maketitle
\begin{abstract}
We report a deep radio search with the Ooty Radio Telescope (ORT) for the 
redshifted 21 cm absorption line from the \dlym system seen at redshift 
3.388 against the quasar PKS 0201+113. This is currently the most distant
system for which a detection of 21 cm absorption has been claimed. The 
present observations have a sensitivity comparable to the earlier ones
and detect no statistically significant absorption. We use the non-detection
to place an upper limit of $\sim$ 0.011 on the optical depth of the \dlym
absorber. This corresponds to a lower limit of $\sim$ 5600 K to the spin 
temperature of the system. This is considerably higher than the previous
upper limit of $\sim$ 1380 K.\\
\end{abstract}
\label{firstpage}
\section{Introduction.}
	In the highest HI column density systems seen in absorption against
distant quasars, the line profile shape is dominated not by motions of the
gas but rather by the intrinsic Lorentzian wings. These so-called damped~\lym 
systems are the dominant known repository of neutral gas at high redshifts.
Systematic optical studies of such systems have established that 
$\Omega_{HI}$, the ratio of the density of neutral gas  to the critical 
density, increases rapidly with increasing redshift and, at high redshift, 
(z $\geq$ 2) is similar to $\Omega_{stars}$, (the ratio of the density of 
luminous material to the critical density), in the local universe 
(Lanzetta et al. 1991, Rao $\&$ Briggs 1995). This is 
consistent with the idea that the \dlym systems  are the 
precursors of $z=0$ disk galaxies.\\
	However, the comoving number density of damped~\lym systems is 
approximately 4 times that of normal spiral galaxies, (Wolfe et al. 1986)
, so, if these systems are truly the precursors 
of spirals, then either there were more precursors than current spirals or 
the precursors had, on the average, much larger disks. Another significant 
difference between the observed properties of \dlym systems and nearby spirals 
is that the inferred spin temperature, $T_s$, of the gas in damped~\lym 
systems is a factor of $\sim 5$ larger than that observed, for similar HI 
column densities, either in our galaxy or Andromeda (Wolfe $\&$ Davis 1979, 
Carilli et al. 1996). The spin temperature is inferred from the 
combination of the optical depth in the 21~cm line and 
the column density, which is in turn obtained from the equivalent width of 
the damped profile. This derivation rests on the assumption that the gas seen 
in absorption against the optical quasars (which have miniscule transverse 
sizes) completely covers the entire radio continuum emitting region of the 
quasars (which are often tens of kiloparsecs in size). Direct VLBI 
observations of 0458-020 (Briggs et al.~1989) have shown that, in this 
one case at least, the absorbing gas has a transverse extent of at least 
8 kpc.\\
	In this paper, we report a deep radio search for the redshifted
21 cm absorption line from the damped \lym system seen at $z = 3.388$ towards 
the quasar PKS 0201+113. This is the highest redshift system for which a 
detection of 21 cm HI absorption has been reported (de Bruyn, O'Dea $\&$ 
Baum 1996, Briggs, Brinks $\&$ Wolfe 1997). de Bruyn et al. used the 
Westerbork Synthesis Radio Telescope (WSRT) 
and reported a peak 21 cm optical depth of $\sim 0.085$ with a very narrow 
velocity width ($\sim 9$ km s$^{-1}$). Briggs et al. observed the system 
with both Arecibo and the VLA. The Arecibo spectrum shows an absorption 
feature with a peak optical depth of $\sim 0.037$, but with a much larger 
velocity extent ($\sim 25$ km s$^{-1}$) than that seen at the WSRT. The 
VLA spectrum showed no statistically significant absorption; this is 
marginally consistent with the Arecibo result, given the sensitivity of 
the two observations. Our own observations have a noise level which is
intermediate between the levels of the Arecibo and VLA spectra and 
also show no statistically significant absorption feature.\\ 
	The rest of the paper is divided as follows; section 2 describes 
the observations, data reduction and results while section 3 contains 
a discussion of the current and previous observations.\\
\vskip 0.2 in
\section{Observations and Results.}
	Ooty Radio Telescope (Swarup et al. 1971) observations of PKS 
0201+113 were carried out in two stretches, the first from the 29th 
of December, 1996 to the 5th of January, 1997 and the second from 
the 14th to the 17th of January, 1997. The source 
was observed for about 8 hours (broken up into several 2 to 2.5 hour 
sessions) each day, with a spectrum being recorded every 2 minutes. 
The observations were performed using the telescope as a two 
element interferometer with the northern half of the telescope being 
correlated with the southern half. The backend bandwidth of 768~kHz was 
divided into 256 spectral channels, giving a frequency resolution of 6 kHz 
after a single Hanning smoothing. The centre of the band was at a sky frequency 
of 323.704~MHz. A strong interference line seen at the edge of the band 
was traced to various PCs at the observing site and was greatly reduced 
by switching most of them off during the observations.\\
         The data reduction was carried out using WASP (Chengalur 1996).
The system gain was calibrated using the radio source Q0710+118, (Bogers 
et al., 1994) whose declination is similar to that of 0201+113; the ORT 
gain is a function of declination. A flux density of 11.1 Jy was assumed
for the purpose of calibration. The calibrator was observed 
for 10 minutes at the end of each observing session; the gain was found
to be stable to within a few percent over our entire observing run. The
raw spectra were corrected for the instrumental gain and bandpass shape
using the calibrator data. The corrected spectra were then carefully 
inspected for any contamination due to interference and a total of 
$\sim$ 17 hours of data was discarded on this basis. An iterative procedure 
was then run, which fitted a third order polynomial to each 2 minute spectrum, 
identified `bad' channels, and then created a summary spectrum of all 
good channels. This was done separately for the data from each 2.5 hour 
session. "Bad" channels were defined as those in which the rms (over time)
exceeded 2 times the rms noise for `good' channels. Further, any individual 
spectral point with an absolute deviation from the median greater than 6 times 
the average rms of "good" channels was also flagged to remove isolated 
radio frequency interference (RFI ) spikes. This procedure removed most 
of the RFI; however, some low level interference did persist at channels 
163-165 and channels 194-195. These regions were blanked in the final 
spectrum before applying the final smoothing.\\
The spectral baseline in an interferometric setup contains contributions
from the visibilities of distant continuum sources, our polynomial 
fitting procedure removes much of this contribution. We also tried
fitting first and second order polynomials to the spectra, the 
results are substantially the same, except for the presence of a low level
broad ripple across the spectrum. Since this ripple and our bandwidth
are both separately much broader than the width of the absorption line 
that we are looking for, our baseline fitting procedure should not
significantly bias our detection limits.\\
	In a small fraction ($\sim$ 5 hours) of the data, extremely strong 
absorption (60-100 mJy ) was seen close to the expected location of the 
redshifted 21~cm line. This feature was, however, extremely narrow ($\sim$ 
1-2 channels wide). The feature was seen for about 1 hour on 4 separate days 
of observation. Since the data stretches immediately before and after these 
1 hour stretches displayed no such absorption, it seems likely that it is 
caused by some kind of interference, although its nature is not clear. 
(Note that the ORT is equatorially mounted; hence its beam does not rotate 
across the sky.) These data stretches were also dropped before creating 
the final averaged spectrum. The latter was created by the weighted 
average of the Doppler shifted 2.5 hour summary spectra, where the per 
channel weights account for both the total integration time in each summary 
spectrum and the number of spectral points which were flagged. The Doppler 
shifting was done by linear interpolation; the total shift across the entire 
observing run was less than one smoothed channel. The final spectrum contains 
the data from 76 hours of observations.\\
        The final edited spectrum is shown in Fig.~1[A] with the expected line
frequency (determined from optical spectroscopy), indicated along with 
its 1$\sigma$ error bars. The rms noise level on the final spectrum is 
$\sim$ 3.7 mJy, which agrees well with the expected noise level. Gaussians 
of various widths were then fitted to the data and the most significant 
feature is found to occur at a depth of $\sim$ 3.7 mJy with a width
of $24 \pm 6$~ kHz and a central frequency of 323.758 $\pm$ 0.006 MHz.
Both the central frequency and the FWHM are similar to the Briggs result 
($\nu_o=323.764 \pm 0.005$ MHz and ${\Delta}{\nu} = 25 \pm 5$ kHz); 
however, the depths of the two features are quite different. Fig.~1[C] shows 
the ORT spectrum smoothed to a resolution of 24 kHz for the purpose
of comparison with the Arecibo result, shown here as a dotted line. 
de Bruyn et al. detected (albeit at low significance) a very narrow line,
their model fit to the line is shown superimposed on the ORT spectrum 
(smoothed to a resolution of 9 kHz to match the WSRT spectrum) in Fig.~1[B].\\
\vskip 0.2 in
\begin{figure*}
\input epsf
\epsfysize = 4.5 in
\epsfxsize = 6.5 in \epsfbox{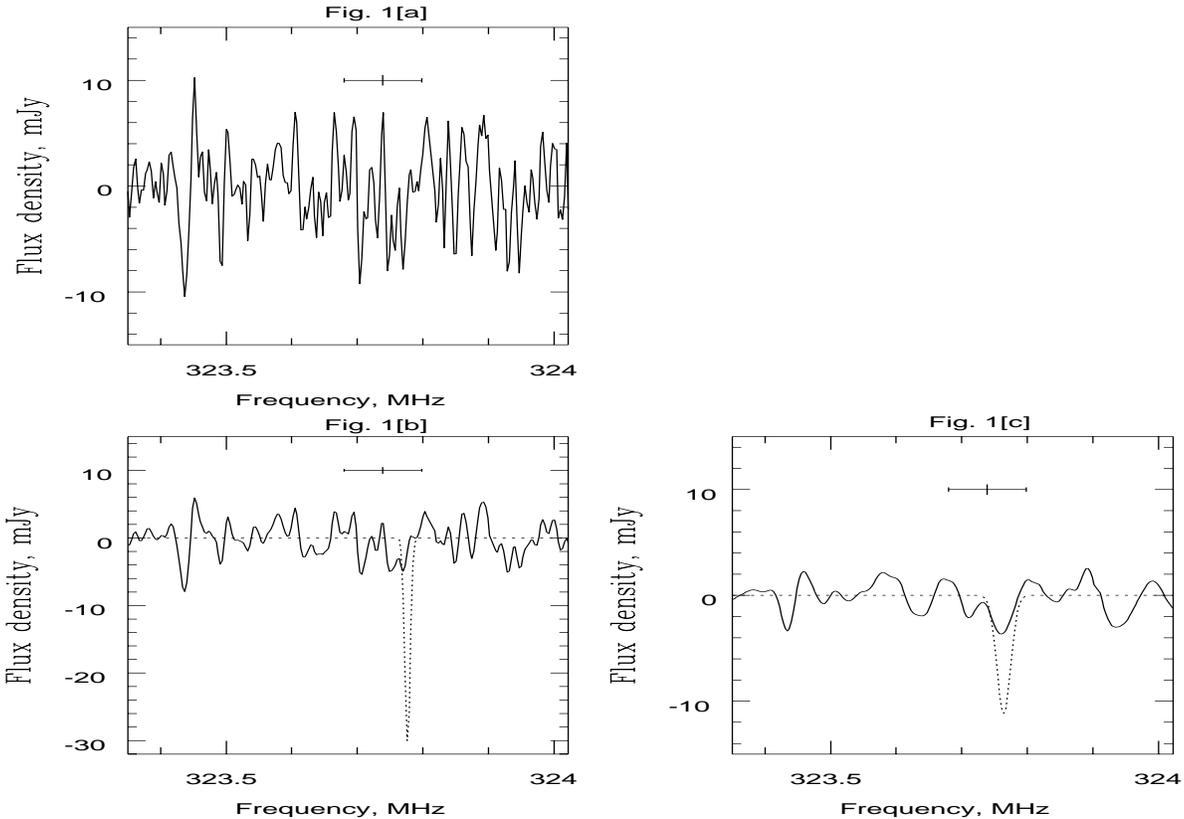}
\caption{ [a] Spectrum of PKS 0201+113. The 21 cm line frequency
corresponding to the optical redshift is shown, along with 1$\sigma$ error 
bars. [b] shows the spectrum smoothed to a resolution of 9 kHz; the 
Gaussian fit to the WSRT feature is shown as a dotted line. [c] shows 
the spectrum smoothed to a resolution of 24 kHz for comparison with the 
Arecibo absorption line, shown here as a dotted line.}
\end{figure*}
\vskip 0.2 in
\section{Discussion.}
	There are two major issues which need to be addressed, (1) the 
implications of our observations, taken by themselves, for the physical 
conditions in the damped~\lym absorber and (2) a comparison between the 
present observations and the earlier ones made with other telescopes. We 
consider each of these in turn.\\
	As discussed in the introduction, radio observations of the 
redshifted 21 cm absorption from damped~\lym systems have traditionally been
used to estimate the spin temperature in these systems. The critical
assumptions involved here are that the damped~\lym system is homogeneous
and completely covers the background radio source. The column density of 
HI obtained along the narrow line of sight to the background quasar can 
then be used in conjunction with the observed optical depth in the 21 cm line
to estimate the spin temperature of the gas using the standard expression 
(eg. Rohlfs 1986)\\ 
$$N_{HI} = 1.82 \times 10^{18}\int{\tau{T_s}dV} $$
where $\tau$ is the optical depth in the 21 cm line; N$_{HI}$ is in per 
cm$^{2}$, T$_s$ in K and dV in km s$^{-1}$.\\
	In the present case, the HI column density is N$_{HI}$ = 2.51 
$\times$ 10$^{21}$ per cm$^2$ (White, Kinney $\&$ Becker 1993). We use a 
continuum flux density of 350 mJy (de Bruyn et al. 1996) and obtain an 
upper limit of ${\tau}_{max}$ = 0.011 to the optical depth (Briggs et al. 
quote a continuum flux density of 290 $\pm$ 5 mJy; using this 
value would give ${\tau}_{max}$ = 0.013). Since the optical depth is 
inversely related to T$_s$, this implies a {\it lower} limit of $\sim$ 
5600 K to the spin temperature.\\
	0201+113 is a compact GPS source (with angular size $<$ 5 mas at 1.6 GHz, 
Hodges, Mutel $\&$ Phillips 1984), making it likely that the damped~\lym 
absorber completely covers the radio continuum. On the other hand, as 
noted by de Bruyn et ~al., the 18 cm VLA maps (Stanghellini et al. 
~1990) show a weak secondary component located $\sim$ 2'' 
(10~kpc at the redshift of the absorber) south of the optical quasar; 
however, the flux in this component (at 18 cm) is only a small fraction 
of the total flux. If the absorbing gas is not all at the same spin 
temperature, then, as has been long appreciated, the derived spin 
temperature is the harmonic mean of the spin temperatures of each 
of the constituents, weighted by their column densities. Our lower limit 
to T$_s$ should be regarded as a lower limit to the phase with the highest 
spin temperature. We note also that our observations have low sensitivity 
to gas with a very large ($\sim$ a few hundred km s$^{-1}$) velocity 
dispersion; the spin temperature calculation assumes that there is very 
little gas with such large dispersion.\\ 
	Turning now to the second issue, viz. that of a comparison between
the present observations and the earlier ones, we begin by noting that our 
lower limit of $\sim$ 5600~K to the spin temperature appears to be at odds 
with the Arecibo and WSRT results of T$_s$ $\sim$ 1380 K and T$_s$ $\sim$ 
1100 K, respectively. (Although the line profiles obtained from the WSRT 
and Arecibo observations are vastly different, the integrated optical depth 
inferred from both these observations are roughly the same; this results in 
similar values for the spin temperature.) Our results are in excellent 
agreement with the VLA spectra in Briggs et al. (1997) and are marginally 
consistent with the Arecibo spectra if one allows the possibility of 3$\sigma$ 
deviations from both results. The WSRT detection is in strong disagreement with 
all the other observations; however, the rms noise level there is much higher 
than that of the others. Both the Arecibo and the WSRT features were seen to 
show the expected Doppler shift over the observing period. The ORT observations 
were carried out $\sim$ 3 years after the Arecibo detection with the VLA 
observations 2 years prior to this. This gives rise to the interesting 
possibility that physical conditions in the source (or, possibly, in the 
absorbing system) might vary on these time scales, thereby causing the 
discrepancy in the results. Such an explanation also removes the necessity for 
fortuitous noise effects which must be otherwise postulated to bring the 3 
observations into marginal consistency. We consider this possibility below.\\
        Initially, we note that the depth of the line is a function both of the 
incident continuum flux density and the optical depth of the absorbing system. 
This implies that changes in the physical conditions in either the source or 
the \dlym cloud can have drastic effects on the line depth. Variability in the 
absorbing system is unlikely due to the difficulty in co-ordinating the 
variation in the properties of a cloud of size of the order of a few parsecs. 
Further, motion of (or within) the cloud itself also seems an implausible 
explanation as it would require relativistic speeds unless the cloud was 
extremely inhomogenous. On the other hand, PKS0201+113 is an active galactic 
nucleus which is extremely compact. Hence, variation of its continuum flux 
over time scales of $\sim$ 2 years would not be extremely surprising. Both 
the ORT and Arecibo observations did not have simultaneous measurements of 
the continuum flux but rely on earlier results. If the Arecibo observations 
were carried out at a time when the flux density was significantly higher 
than the VLA value, a high value of $\tau$ would be estimated which would 
result in a low estimate of the spin temperature. We note that the VLA and 
WSRT estimates of the quasar flux differ by $\sim$ 20 per cent over a period 
of $\sim$ 6 months. However, even if this difference is interpreted as being 
entirely due to variablilty of the source, it is not quite enough to produce 
the observed differences in the spectra; a change in the flux density by 
a factor of $\sim$ 2 is required, on a time scale of $\sim$ 2 years. This 
is, of course, presently in the realm of speculation and the source should 
probably be monitored to confirm or deny its variability before any firm 
conclusions can be drawn about the physical conditions in the system.\\
\vskip 0.2 in
\section{Summary.}
We report ORT observations of the \dlym system at z = 3.388 towards the quasar 
PKS0201+113. No absorption feature deeper than 3.7 mJy is seen at or near the 
expected location of the 21 cm line. This places an upper limit of 0.011 on the 
optical depth, $\tau$; earlier estimates of HI column density from \dlym 
observations are then used to derive a lower bound of $\sim$ 5600 K on the spin 
temperature of the absorbing system. This is considerably larger than the previous 
upper limit of $\sim$ $1380$~K.
\vskip 0.2 in
\section*{Acknowledgments.}
We are immensely grateful to D. Anish Roshi, without whose help these 
observations would not have been possible. Ger de Bruyn and Frank Briggs 
provided useful comments on preliminary results; we thank them for this. 
We are also grateful to the observing staff of the ORT, particularly 
Bilal and Magesh, for their assistance in carrying out these somewhat 
non-standard observations. Finally, we thank Gopal-Krishna for a critical
reading of an earlier draft of this paper.\\
\vskip 0.2 in
\section*{References.}
Bojers, W.J., Hes, R., Barthel, P.D. $\&$ Zensus, J.A., 1994, A$\&$AS, 105, 91\\
Briggs, F.H., Wolfe,A.M., Liszt, H.S., Davis, M.M. $\&$ Turner, K.L., 1989, 
ApJ, 341, 650\\
Briggs, F.H., Brinks, E. $\&$ Wolfe, A.M., 1997, AJ, 113, 467\\
Carilli, C.L.,Lane, W., de Bruyn, A.G., Braun, R. $\&$ Miley, G.K ., 1996, 
AJ, 111, 1830 \\
Chengalur, J.N., 1996, NFRA Note 453\\
de Bruyn, A.G., O'Dea, C.P. $\&$ Baum, S.A., 1996, A$\&$A, 305, 450\\
Hodges, M.W., Mutel, R.L. $\&$ Phillips, R.B., 1984, AJ, 89, 1327\\
Lanzetta, K.M., Wolfe, A.M., Turnshek, D.A. Lu, L., McMahon, R.G. $\&$ Hazard, 
C., 1991, ApJS, 77,1\\
Rao, S.A. $\&$ Briggs, F.H., 1995, ApJ, 449, 488\\
Rohlfs, K., 1986, Tools of Radio Astronomy, (Springer-Verlag, Berlin Heidelberg)\\
Stanghellini, C., Baum, S.A., O'Dea, C.P. $\&$ Morris, G.B., 1990, A$\&$A, 233, 379\\
Swarup, G. et al., 1971, Nature Physical Sciences, 230, 185\\
White, R.L., Kinney, A.L. $\&$ Becker, R.H., 1993, ApJ, 407, 456\\
Wolfe, A.M. $\&$ Davis, M.M., 1979, AJ, 84, 699\\
Wolfe, A.M., Turnshek, D.A., Smith, H.E. $\&$ Morris, G.B.., 1986, ApJS, 61, 249\\
\vskip 0.2 in
\bsp
\label{lastpage}
\end{document}